\documentclass[final,5p,times]{elsarticle}
\usepackage[T1]{fontenc}
\usepackage[utf8]{inputenc}
\usepackage{xcolor}
\usepackage{amsmath}
\usepackage{amsthm}
\usepackage{amssymb}
\usepackage{graphicx}
\usepackage[bookmarks=false,
 breaklinks=false,pdfborder={0 0 1},backref=section,colorlinks=true]
 {hyperref}
\hypersetup{
 citecolor=blue,linkcolor=blue,urlcolor=blue}

\makeatletter
\theoremstyle{plain}
\newtheorem{thm}{\protect\theoremname}

\usepackage{graphicx}\usepackage{bm}

\makeatother

\providecommand{\theoremname}{Theorem}

\begin{document}
\title{Spectral mechanism and nearly reducible transfer matrices for pseudotransitions
in one-dimensional systems}
\author{Onofre Rojas}
\address{Department of Physics, Institute of Natural Science, Federal University
of Lavras, Lavras, MG, Brazil}
\begin{abstract}
While true phase transitions are forbidden in one-dimensional systems
with short-range interactions, several models have recently been shown
to exhibit sharp yet analytic thermodynamic anomalies that mimic thermal
phase transitions. We show that this behavior arises from transfer
matrices that are mathematically irreducible but possess a nearly
block-diagonal structure due to the weak contribution of off-diagonal
Boltzmann weights in the low-temperature regime. This results in weakly
coupled competing sectors whose eigenvalue competition produces abrupt
crossovers without nonanalyticity, a mechanism we term nearly block-diagonal
irreducible. A key thermodynamic signature of such pseudotransitions
is that the residual entropy at the interface remains bounded between
the residual entropies of the competing sectors. We develop a general
spectral framework to describe this behavior and apply it to two representative
models: the Ising chain with internal degeneracy (Doniach model) and
a hexagonal nanowire chain with mixed spin-1/2 and spin-1 components.
In the first case, we derive exact expressions for the pseudo-critical
temperature and residual entropy. In the second, we reduce the full
$1458\times1458$ transfer matrix via symmetry decomposition and construct
a low-rank effective matrix that accurately captures the crossover
between quasi-ferromagnetic and quasi-core-ferromagnetic regimes.
Our results demonstrate that pseudotransitions can be understood as
spectral phenomena emerging from irreducible but functionally decoupled
structures within the transfer matrix.
\end{abstract}
\begin{keyword}
Pseudotransitions; Avoided level crossing; Residual entropy; Nearly
reducible matrices
\end{keyword}
\maketitle

\section{Introduction}
\sloppy
The absence of thermodynamic phase transitions in one-dimensional
(1D) systems with short-range interactions is well established through
rigorous results, including the van Hove theorem \citep{vanhove1950,ruelle}
and its generalization by Cuesta and Sánchez \citep{cuesta}. These
theorems demonstrate the analyticity of the free energy under conditions
of homogeneity and finite-range interactions. The argument relies
on properties of the transfer matrix, which, under the Perron-Frobenius
theorem \citep{ky-lin,ninio,Lavis-2015}, must be irreducible and
strictly positive. This ensures a unique, non-degenerate dominant
eigenvalue, thereby precluding non-analytic behavior in thermodynamic
potentials \citep{cuesta,Lavis-2015}.

Despite this prohibition, a growing body of work over the past decade
has reported thermodynamic anomalies in 1D models that closely resemble
true phase transitions. These features include sharp but continuous
changes in entropy and internal energy, along with pronounced peaks
in specific heat or susceptibility near certain temperatures. Although
the free energy remains analytic, the resulting behavior strongly
mimics both first- and second-order transitions, raising questions
about the origin and interpretation of such anomalies \citep{Isaac-2,Galisova,hutak}.

These anomalies tend to occur at specific temperatures and closely
resemble finite-size effects seen in simulations, such as the sharp
peaks in observables from Monte Carlo studies on finite lattices \citep{psd-MCs}.
This similarity has led to the adoption of the term pseudotransition
to describe such behavior. Timonin \textit{\emph{\citep{Timonin}}},
in his study of spin ice under an external field, used similar terminology
to refer to these sharp anomalies and introduced the concept of quasi-phases
to describe the regions where they arise. Since then, the term has
gained traction in describing these 1D anomalies \textit{\emph{\citep{pseudo}}},
although this feature is sometimes referred to as an ultranarrow phase
crossover or a forbidden phase transition \textit{\emph{\citep{w-yin-prl,w-yin-prb,w-yin-prr,w-yin-arx,Sznajd,Sznajd22}}}.

One important line of investigation concerns the apparent universality
of these anomalies. In several models, they have been characterized
by a consistent set of pseudo-critical exponents: the specific heat
scales with $\alpha=3$, the correlation length with $\nu=1$, and
the susceptibility with $\gamma=3$, suggesting a degree of universality
in their thermodynamic behavior \citep{unv-cr-exp}. Ref.\,\citep{finite-chain}
further supports this picture by analyzing finite-size effects near
the anomaly, showing how the observed behavior sharpens with system
size. Other studies have explored the role of residual entropy near
the anomalous region, revealing additional thermodynamic signatures
\citep{ph-bd}. Recent analyses of diluted Ising chains under constrained
thermodynamic potentials, such as fixed density, reveal that pseudo-critical
exponents and entropy-based criteria are not preserved \citep{yasinskaya},
although the anomalies remain evident.

Correlation functions are central to understanding pseudotransitions
in one-dimensional spin systems, as they reveal the emergence of extended
short-range order despite the absence of true criticality. In exactly
solvable models such as the Ising-Heisenberg and Ising-XYZ diamond
chains, the correlation length exhibits a sharp but finite peak at
the pseudo-critical temperature\citep{Isaac,krokhmalskii2021}. Similar
behavior has been observed in $q$-state Potts-like chains \citep{panov2021}.
This phenomenon arises from a near-degeneracy of subleading eigenvalues
in the transfer matrix, leading to a bifurcation in the correlation
length a key spectral signature of pseudotransitions \citep{chapman2024}.
The resulting slow decay of correlations reflects a large but finite
correlation length, reinforcing the view of these anomalies as remnants
of forbidden phase transitions in low dimensions.

Over the past decade, numerous models have been shown to exhibit such
behavior. These include the Ising chain with internal degeneracy \citep{tiago},
the Ising diamond chain\citep{strecka-book,strecka-dmd}, and more
elaborate systems like the spin-1/2 Ising-Heisenberg diamond chain
and its variants, which display pseudotransitions near ground-state
boundaries \citep{unv-cr-exp,krokhmalskii2021,on-strk}. The diluted
Ising chain studied by Panov links the anomaly to residual entropy
at critical interfaces \citep{yasinskaya,yasinskaya-X,panov2022}.
Other examples include double-tetrahedral chains with mixed spins
in external fields \citep{cond-matt-pecul}, as well as triangular
and ladder Ising-Heisenberg models \citep{chapman2024,on-strk,strk-cav},
many of which can be mapped to simpler effective systems via decoration
transformation. In most cases, the effective transfer matrices reduce
to sizes such as $2\times2$ \citep{pseudo} or $4\times4$, as in
the extended Hubbard chain in atomic limit \citep{hubbard2021,hubbard2024}.
More complex cases include a $6\times6$ matrix in spin-pseudospin
cuprate models \citep{katarina}, an $8\times8$ matrix in the Toblerone-like
Ising chain \citep{chapman2024}, and a $92\times92$ irreducible
matrix in the Blume-Capel nanowire model \citep{Rodrigo}, as well
as general $q\times q$ matrices in Potts-like chains \citep{panov2021,panov2023}. 

Despite these advances, most analyses describe the
sharp crossovers without explaining their origin. Here we identify
the generic two-sector spectral structure that produces pseudotransitions,
derive the corresponding pseudo-critical condition, and show how an
interface residual entropy arises from the constrained low-temperature
limit along the pseudo-critical line. This yields a unified perspective
on the mechanism behind the anomalous behavior reported in many one-dimensional
models.

In this work, we propose a unified framework to understand the origin
of pseudotransitions in one-dimensional systems through the lens of
transfer matrix structure. Sec.\ref{sec:2} motivates the phenomenon
from known examples and ensemble sensitivity. Sec.\ref{sec:3} develops
the central spectral theory, identifying nearly reducible matrices
as the key mechanism behind sharp but analytic crossovers. This includes
explicit spectral criteria, correlation length scaling, and thermodynamic
implications. Sec.\ref{sec:4} and\,\ref{sec:5} apply this framework
to two representative models: an Ising chain with internal degeneracy
(Doniach model\citep{Doniach}) and a mixed-spin nanowire system,
respectively. We conclude in Sec.\ref{sec:6} by discussing thermodynamic
signatures and directions for future generalizations.

\section{Transfer matrix and matrix structure}\label{sec:2}

The transfer matrix technique in statistical physics was introduced
in 1941 by H. Kramers and G. Wannier\citep{kramers}. Since then,
it has been widely used to compute the partition functions of several
lattice models. In this approach, the partition function is expressed
as a sum over all microstates, with each term weighted by the Boltzmann
factor corresponding to the energy of that state.

\subsection{Transfer matrix and irreducibility}

Consider a general non-symmetric transfer matrix $\mathbf{V}\in\mathbb{R}^{n\times n}$
with non-negative entries $v_{i,j}\geqslant0$, written as 
\begin{equation}
\mathbf{V}=\begin{pmatrix}v_{1,1} & v_{1,2} & \cdots & v_{1,n}\\
v_{2,1} & v_{2,2} & \cdots & v_{2,n}\\
\vdots & \vdots & \ddots & \vdots\\
v_{2,1} & v_{2,n} & \cdots & v_{n,n}
\end{pmatrix}.
\end{equation}

A key mathematical concept is the irreducibility of the transfer matrix.
A non-negative matrix $\mathbf{V}$ is said to be irreducible if it
cannot be brought into block upper-triangular form via simultaneous
permutations of rows and columns. Physically, irreducibility ensures
full connectivity between microscopic configurations across the chain.

In order to analyze the eigenvalues of the matrix $\mathbf{V}$, we
use the Perron-Frobenius theorem\citep{ky-lin,ninio,Lavis-2015}.
\begin{thm}
\emph{{[}Perron-Frobenius{]} }Let $\mathbf{V}\in\mathbb{R}^{n\times n}$
be a non-negative, irreducible matrix. Then: 
\begin{itemize}
\item There exists a unique largest eigenvalue $\lambda_{1}>0$, which is
simple (non-degenerate);;
\item the corresponding right eigenvector $|\psi\rangle$has strictly positive
components, i.e., $\langle i|\psi\rangle>0$ for all $i=1,\dots,n$;
\item and all other eigenvalues $\lambda_{j}$ satisfy $|\lambda_{j}|<\lambda_{1}$,
$j=2,\cdots,n$.
\end{itemize}
\end{thm}

In statistical mechanics, this theorem guarantees that in the thermodynamic
limit, the partition function is dominated by a single, positive eigenvalue
$\lambda_{1}$, which controls the free energy per site. This ensures
the system has a well-defined thermodynamic behavior without degeneracies.

However, to fully exclude phase transitions, one must also ensure
that the leading eigenvalue $\lambda_{1}$ varies smoothly with temperature
or other control parameters. This leads to the second essential result:
Kato’s analytic perturbation theorem\citep{kato}.
\begin{thm}
\emph{{[}Kato, adapted{]}} Let \textup{$\mathbf{V}(z)$} be a complex
matrix whose entries are analytic functions of a parameter $z$. If
$\lambda_{1}(z_{0})$ is an isolated, simple eigenvalue, then there
exists an analytic function $\lambda_{1}(z)$ defined near $z_{0}$,
such that both $\lambda_{1}(z)$ and its corresponding eigenvectors
vary analytically with $z$.
\end{thm}

In physical applications, the parameter $z$ often represents temperature
$T$, but it may also correspond to other thermodynamic control variables
such as magnetic field or chemical potential. If the transfer matrix
$\mathbf{V}(z)$ depends analytically on $z$, and its dominant eigenvalue
$\lambda_{1}(z)$ remains simple and isolated, then Kato’s theorem
\citep{kato} ensures that $\lambda_{1}(z)$ and its associated eigenvectors
vary analytically with $z$. In particular, for $z=T>0$, analyticity
holds because Boltzmann weights are well-defined, and this smooth
behavior of the dominant eigenvalue plays a central role in determining
the thermodynamic properties of the system.

Combining these ideas leads to a rigorous conclusion about one-dimensional
systems with short-range interactions, as formalized by Cuesta and
Sánchez\citep{cuesta}.
\begin{thm}
\emph{{[}Non-existence of phase transitions in 1D systems{]}} Let
$\mathbf{V}(x,T)$ be a real, non-negative, irreducible matrix whose
entries are analytic in both a control parameter $x$ and temperature
$T>0$. Let $\lambda_{1}(x,T)$ denote its largest eigenvalue, assumed
to be simple and isolated for all $T>0$. Then the free energy per
site,
\begin{equation}
f(x,T)=-T\ln\lambda_{1}(x,T),
\end{equation}
is analytic in both $x$ and $T$.
\end{thm}

Here and throughout, we set Boltzmann’s constant $k_{B}=1$, so that
$\beta=1/T$ and temperature and inverse temperature may be used interchangeably.

Combining the analytic dependence of the transfer matrix $\mathbf{V}(x,T)$
with the simplicity and isolation of its dominant eigenvalue $\lambda_{1}(x,T)$,
one obtains a rigorous constraint on the thermodynamic behavior of
one-dimensional systems. The result, due to Cuesta and Sánchez \citep{cuesta},
states that the free energy per site, defined in the thermodynamic
limit by the leading eigenvalue, is an analytic function of both temperature
and external parameters, provided $\mathbf{V}(x,T)$ remains irreducible
and analytic. This analyticity rules out the possibility of genuine
phase transitions, as no singularities can arise in $f(x,T)$ under
these conditions.

\subsection{Block-diagonal reducible matrices}\label{subsec:2-a}

To illustrate a scenario that permits non-analytic behavior, let us
consider a transfer matrix $\mathbf{V}$ that is reducible. For simplicity,
assume $\mathbf{V}$ decomposes into two decoupled irreducible submatrices
$\mathbf{A}$ and $\mathbf{B}$, such that
\begin{equation}
\mathbf{V}=\begin{pmatrix}\mathbf{A} & \boldsymbol{0}\\
\boldsymbol{0} & \mathbf{B}
\end{pmatrix}.\label{eq:V-block}
\end{equation}
A prominent example is the Kittel model \citep{kittel},
in which the local constraints forbid transitions between closed and
open configurations, so the transfer matrix remains reducible at all
finite temperatures. Similar behavior appears in $q$-state models
such as Refs. \citep{panov2021,badasyan} when the limit of infinite
$q$ is taken. 

Each block contributes its own dominant eigenvalue, denoted $w_{a}=\langle\phi_{a}|\mathbf{A}|\psi_{a}\rangle$
and $w_{b}=\langle\phi_{b}|\mathbf{B}|\psi_{b}\rangle$, with corresponding
right (left) eigenvectors $|\psi_{a}\rangle$ ($\langle\phi_{a}|$)
and $|\psi_{b}\rangle$ ($\langle\phi_{b}|$), respectively. According
to the Perron-Frobenius theorem, both $w_{a}$ and $w_{b}$ are strictly
positive and non-degenerate within their respective blocks.

The spectrum of $\mathbf{V}$ is simply the union of the spectra of
$\mathbf{A}$ and $\mathbf{B}$. Hence, the two leading eigenvalues
of $\mathbf{V}$ are
\begin{alignat}{1}
\lambda_{1}= & \max\left(w_{a},w_{b}\right),\\
\lambda_{2}= & \min\left(w_{a},w_{b}\right).
\end{alignat}
Suppose now that $\mathbf{V}$ depends on some external parameter
$x$, such as a magnetic field or chemical potential. For a particular
point ($x_{p},T_{p}$), the two leading eigenvalues may coincide,
$w_{a}(x_{p},T_{p})=w_{b}(x_{p},T_{p})$, resulting in a true level
crossing and a degeneracy in the spectrum. The corresponding free
energy per site then becomes
\begin{equation}
f=-T\ln\left[\max\left(w_{a},w_{b}\right)\right],\label{eq:free-dsc}
\end{equation}
which is non-analytic at the crossing point $(x_{p},T_{p})$, signaling
a first-order phase transition between decoupled sectors $a$ and
$b$. 

At first glance, this seems to contradict the non-existence theorem
discussed earlier. However, the key assumption of irreducibility is
no longer satisfied: the decoupling of $\mathbf{A}$ and $\mathbf{B}$
violates the conditions of both the Perron-Frobenius and Cuesta-Sánchez
theorems. The resulting non-analyticity is therefore consistent with
the underlying theory.

This simple two-block decomposition captures the mechanism
by which strict reducibility allows genuine thermodynamic singularities.
The Kittel model\citep{kittel} is a representative example of this
situation. We return to this point in Sec. \ref{subsec:3-a}, where
the comparison with nearly reducible but fully connected matrices
clarifies why pseudotransitions arise even though the transfer matrix
remains irreducible.

\subsection{Nearly block-diagonal irreducible matrices}\label{subsec:3-a}

Let us now consider a more subtle case: the transfer matrix $\mathbf{V}$
is formally irreducible, but exhibits a nearly block-diagonal structure
\begin{equation}
\mathbf{V}=\begin{pmatrix}\mathbf{A} & \mathbf{C}\\
\mathbf{D} & \mathbf{B}
\end{pmatrix},
\end{equation}
where $\mathbf{A}$ and $\mathbf{B}$ are dominant irreducible blocks, while
 $\mathbf{C}$ and $\mathbf{D}$ contain small off-diagonal couplings. Although
$\mathbf{V}$ is formally irreducible, the elements of $\mathbf{C}$ and $\mathbf{D}$ 
may become suppressed at a given parameter value $x_{p}$, nearly
decoupling the sectors $\mathbf{A}$ and $\mathbf{B}$ at a fixed
temperature.

To analyze the thermodynamics in this limit, we restrict $\mathbf{V}$
to the subspace spanned by the leading eigenvectors $|\psi_{a}\rangle$,
$|\psi_{b}\rangle$ of A and B, respectively. This yields the effective
two-level matrix
\begin{equation}
\mathcal{V}_{t}=\begin{pmatrix}w_{a} & \langle\phi_{a}|\mathbf{C}|\psi_{b}\rangle\\
\langle\phi_{b}|\mathbf{D}|\psi_{a}\rangle & w_{b}
\end{pmatrix},\label{eq:V-eff}
\end{equation}
where $\langle\phi_{a}|\mathbf{C}|\psi_{b}\rangle>0$ and $\langle\phi_{b}|\mathbf{D}|\psi_{a}\rangle>0$,
according to the condition of matrix non-negative and irreducibility,
then let us define $\langle\phi_{b}|\mathbf{D}|\psi_{a}\rangle\langle\phi_{a}|\mathbf{C}|\psi_{b}\rangle=w_{0}^{2}$. 

The emergence of an effective two-sector description
follows from the low-temperature spectral structure of the transfer
matrix. In strictly reducible cases such as the Kittel model \citep{cuesta,kittel},
the dominant eigenvalues $w_{a}$ and $w_{b}$ belong to disconnected
blocks. In the irreducible but nearly block-structured situation considered
here, these quantities play the role of diagonal statistical weights
for two competing low-energy configurations, while the off-diagonal
term $w_{0}$ represents the minimal excitation linking them. Near
a pseudotransition the dominant eigenvalues lie exponentially closer
to each other than to the rest of the spectrum, so subdominant eigenvectors
decay on much shorter length scales and remain almost orthogonal to
the space spanned by $|\psi_{a}\rangle$ and $|\psi_{b}\rangle$.
Since $w_{0}$ is also exponentially small, hybridization with higher
sectors is negligible, and the thermodynamics is effectively governed
by the resulting $2\times2$ structure.

Thus, the eigenvalues of $\mathcal{V}_{t}$ are:
\begin{equation}
\lambda_{1,2}=\tfrac{1}{2}\Bigl(w_{a}+w_{b}\pm\sqrt{(w_{a}-w_{b})^{2}+4w_{0}^{2}}\Bigr),\label{eq:lamb_12}
\end{equation}
which exhibit an avoided crossing due to finite $w_{0}$ even when
$w_{a}\approx w_{b}$. This guarantees analyticity of the free energy,
despite the presence of a sharp thermodynamic crossover. 

To make this structure more transparent, define the average eigenvalue
$\bar{w}=\frac{1}{2}(w_{a}+w_{b})$, and introduce the dimensionless
spectral splitting: 
\begin{alignat}{1}
\zeta= & \frac{\sqrt{\sigma^{2}+4w_{0}^{2}}}{\bar{w}},\label{eq:zeta-def}
\end{alignat}
with $\sigma=w_{a}-w_{b}$. The eigenvalues can then be rewritten
as
\begin{alignat}{1}
\lambda_{1,2}= & \bar{w}\left(1\pm\tfrac{1}{2}\zeta\right).\label{eq:L12}
\end{alignat}

We consider this case to show how weak inter-block couplings, though
negligible at low temperatures, can lift degeneracies and induce avoided
crossings between dominant eigenvalues at finite temperatures.

Thus, the free energy per site in the thermodynamic limit becomes
analytic and is given by
\begin{equation}
f=-\tfrac{1}{\beta}\ln\left[\bar{w}\left(1+\tfrac{1}{2}\zeta\right)\right].\label{eq:free-energ}
\end{equation}

As $w_{0}\to0$, the $\zeta$ function \eqref{eq:zeta-def} decreases
and the spectral gap narrows, producing a sharp but analytic crossover.
In this limit, the off-diagonal couplings vanish, and the pseudotransition
sharpens toward an effective level crossing. 

\subsubsection{Avoided Crossing and Minimal Spectral Gap}

\begin{figure}
\includegraphics[scale=0.6]{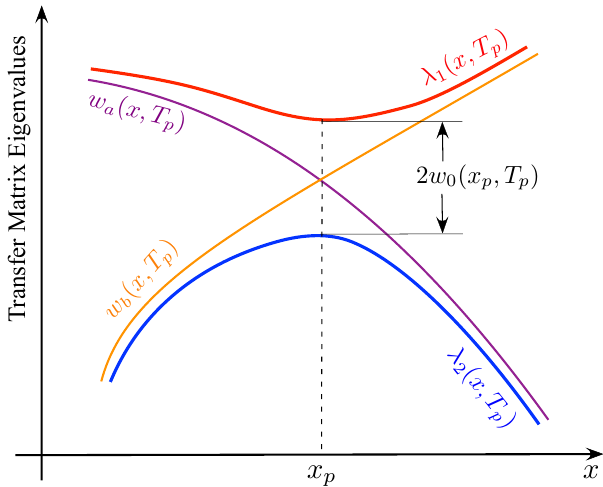}\caption{Schematic representation of the two leading eigenvalues$\lambda_{1}$
and $\lambda_{2}$ as a function of control parameter $x$ at fixed
temperature $T_{p}$. The minimal gap $2w_{0}$ appears at $x=x_{p}$,
where $w_{a}=w_{b}$ when $w_{0}\to0$. }\label{fig:Schematic-representation-of}
\end{figure}

To visualize the analytic structure discussed above, consider the
behavior of the dominant eigenvalues $\lambda_{1}(x,T_{p})$ and $\lambda_{2}(x,T_{p})$
near the pseudo-critical point $(x_{p},T_{p})$. As illustrated in
Fig.\,\ref{fig:Schematic-representation-of}, these eigenvalues exhibit
an avoided crossing: when the coupling $w_{0}\to0$, the weights $w_{a}$
and $w_{b}$ become degenerate at $x=x_{p}$, and the minimal gap
between $\lambda_{1}$ and $\lambda_{2}$ closes asymptotically.

The leading-order condition for this crossover is given by the spectral
balance $w_{a}(x_{p},T_{p})=w_{b}(x_{p},T_{p})$, which identifies
the point $(x_{p},T_{p})$ where the dominant sectors have equal statistical
weight. This condition underlies many pseudotransition studies \citep{pseudo,w-yin-prl,w-yin-prb,w-yin-prr,krokhmalskii2021,panov2021,chapman2024,katarina}
and will be revisited in Sec. \ref{subsec:Psd-crit-cond} with explicit
thermodynamic consequences.

The $\zeta$ function also can be approximate around the avoiding
crossing through eigenvalue ratio:
\begin{equation}
\left(\frac{\lambda_{2}}{\lambda_{1}}\right)=\left(\frac{1-\zeta/2}{1+\zeta/2}\right)\approx{\rm e}^{-\zeta},
\end{equation}
or, equivalently, eq.\eqref{eq:zeta-def} can be more conveniently
written as 
\begin{equation}
\zeta=\ln\left(\frac{\lambda_{1}}{\lambda_{2}}\right).\label{eq:zeta-lambd}
\end{equation}

This logarithmic form directly relates the spectral gap to the correlation
length, which becomes $\xi=\zeta^{-1}$ in the absence of competing
subdominant eigenvalues. 

With the phenomenological context in place, we now turn to the central
spectral analysis. Sec.\ref{sec:3} introduces the general formalism
of nearly reducible transfer matrices and their thermodynamic implications.
This section forms the theoretical core of the paper and will serve
as the reference point for all model-specific analyses that follow.

\section{Thermodynamic consequences of nearly block-diagonal structure}\label{sec:3}

As discussed in Sec. \ref{subsec:2-a}, small off-diagonal elements
in an otherwise block-structured transfer matrix prevent true level
crossings and enforce analyticity. Nevertheless, when the dominant
sectors are nearly degenerate, these weak couplings can generate sharp
thermodynamic crossovers that closely resemble phase transitions.
In this section, we examine the thermodynamic implications of this
structure.

\subsection{Expansion of the spectral gap and pseudo-critical point}

To analyze the thermodynamic behavior close to a
pseudotransition, we expand the normalized spectral gap $\zeta(x,T)$
in a Taylor series around the anomalous point $(x_{p},T_{p})$. Analyticity
of $\zeta$ in both variables is guaranteed for any finite irreducible
transfer matrix by Kato’s theorem \citep{kato}, and the presence
of the mixing term $w_{0}>0$ ensures that $\zeta$ remains strictly
positive even when $w_{a}=w_{b}$. The detailed expressions for the
derivatives, including mixed terms, are derived in Appendix \ref{subsec:Appendix-A}.

For small deviations $\delta T=T-T_{p}$ and $\delta x=x-x_{p}$,
the spectral gap admits the local expansion
\begin{alignat}{1}
\zeta(x,T)= & \zeta(x_{p,}T_{p})+\zeta_{T_{p}}\delta T+\zeta_{x_{p}}\delta x+\nonumber \\
 & +\zeta_{T_{p}^{2}}\frac{\delta T{}^{2}}{2}+\zeta_{x_{p}^{2}}\frac{\delta x{}^{2}}{2}+\zeta_{x_{p},T_{p}}\:\delta x\,\delta T\cdots,
\end{alignat}
where all derivatives ($\zeta_{T}=\frac{\partial\zeta}{\partial T}$,
$\zeta_{x}=\frac{\partial\zeta}{\partial x}$, $\zeta_{T^{2}}=\frac{\partial^{2}\zeta}{\partial T^{2}}$,
$\zeta_{xx}=\frac{\partial^{2}\zeta}{\partial x^{2}}$, $\zeta_{x,T}=\frac{\partial^{2}\zeta}{\partial x\partial T}$)
are evaluated at $(x_{p},T_{p})$. This expansion is valid within
the narrow region where the pseudo-critical crossover occurs, that
is, for $|\delta T|$ and $|\delta x|$ small enough that higher-order
terms remain negligible. The mixed derivative $\zeta_{xT}$ quantifies
how variations in the control parameter and temperature jointly affect
the curvature of the gap valley, determining the tilt of the minimum
in the $(x,T)$ plane and how its position shifts when $x$ and $T$
vary simultaneously.

Symbolically, the derivative of $\zeta$ with respect to a generic
parameter (such as $x$ or $T$) can be expressed in terms of $\bar{w}$,
$\sigma$, and $w_{0}$ as follows

\begin{equation}
\zeta'=\frac{\left(\sigma\sigma'+4w_{0}w_{0}'\right)\bar{w}-(\sigma^{2}+4w_{0}^{2})\bar{w}'}{\bar{w}^{2}\sqrt{\sigma^{2}+4w_{0}^{2}}}.\label{eq:zeta'}
\end{equation}

Here, the prime notation denotes the derivative with respect to a
general parameter, either $x$ or $T$, depending on the context.
Fixing $T_{p}$, $\zeta$ attains a minimum at $x_{p}$ if
\begin{alignat}{1}
\zeta_{x_{p}}= & \frac{\partial\zeta(x,T_{p})}{\partial x}|_{x_{p},T_{p}}=0.\label{eq:cond-x_p}
\end{alignat}
Similarly, we can evaluate the partial derivative $\zeta_{T}$ at
the point $(x_{p},T_{p})$, yielding 

\begin{alignat}{1}
\zeta_{T_{p}}= & \frac{\partial\zeta(x_{p},T)}{\partial T}|_{x_{p},T_{p}}\ne0,\label{eq:cond-Tp}
\end{alignat}

This condition characterizes a valley structure in the $x-T$ plane,
where the depth or height of the valley is determined by the value
of $\zeta(x,T)$. In general, the minimum does not correspond to a
well (i.e., a true critical point), which would require the gradient
to vanish in both directions: $\nabla\zeta=(0,0)$ at $(x_{p},T_{p})$.
Instead, the valley indicates a sharp but analytic crossover consistent
with a pseudotransition.

On the other hand, by fixing $x_{p}$, we can minimize $\zeta$ with
respect to temperature. This yields the condition:

\begin{alignat}{1}
\zeta_{T_{p^{*}}}= & \frac{\partial\zeta(x_{p},T_{p^{*}})}{\partial T}|_{x_{p},T_{p^{*}}}=0,\label{eq:cond-T_p*}
\end{alignat}
for some temperature $T_{p^{*}}$. This defines the point where the
spectral gap reaches its minimum along the temperature axis at fixed
$x_{p}$.

Notably, when the difference $\Delta T=T_{p}-T_{p^{*}}\rightarrow0$,
the valley in the $x-T$ landscape becomes sharply localized, resembling
a narrow canyon, as illustrated in Fig.\,\ref{fig:Surface-of-zeta}.
This geometric narrowing of the gap region reflects the robustness
of the pseudotransition and marks the regime where thermodynamic quantities
exhibit sharp but continuous changes.

\subsubsection{Pseudo-critical temperature}\label{subsec:Psd-crit-cond}

In the regime where the off-diagonal coupling is small $(0<w_{0}\ll w_{a}\sim w_{b}$),
the condition for the pseudo-critical temperature follows from the
general minimum conditions for the spectral gap $\zeta(x,T)$, as
derived in Appendix \ref{subsec:Appendix-A}. In particular, Eq.\,\eqref{eq:gen-Tp*}
gives the temperature minimization of $\zeta$, and Eq.\,\eqref{eq:gen-xp}
provides control parameter minimization. In the zeroth-order limit
where excited-state contributions are negligible and $w_{0}\to0$,
these lead directly to the simplified \textsl{balance condition}
\begin{equation}
w_{a}(x_{p},T_{p})=w_{b}(x_{p},T_{p}),\label{eq:w1-w2}
\end{equation}
 discussed earlier. This expression captures the leading-order estimate
for the pseudo-critical point where the statistical weights of the
dominant sectors become equal. This is the well known condition explored
in the literature when discussed the pseudo-critical temperature or
ultra-narrow crossover point\citep{pseudo,w-yin-prl,w-yin-prb,w-yin-prr,krokhmalskii2021,panov2021,chapman2024,katarina}.

Physically, the condition $w_{a}=w_{b}$ implies
equality of the free energies of two sectors, kind of true phase transition
when ignoring $w_{0}$ contribution. However, when $w_{a}\approx w_{b}$,
the two leading eigenvalues \eqref{eq:L12} is out of the minimum
and sharp changes vanishes.

A pseudotransition occurs when the dominant eigenvalues $w_{a}$ and
$w_{b}$ of two weakly coupled sectors become nearly degenerate. Assuming
the off-diagonal coupling $w_{0}$ is small, the pseudo-critical temperature
$T_{p}$ can be estimated using Eq.\, \eqref{eq:w1-w2}, derived in
Appendix\, \ref{subsec:Appendix-A}. Retaining only the ground and
first excited states in each sector, one obtains:
\begin{equation}
e^{-\epsilon_{p}/T_{p}}=\frac{g_{b,0}+g_{b,1}e^{-\delta_{b,1}/T_{p}}}{g_{a,0}+g_{a,1}e^{-\delta_{a,1}/T_{p}}},\label{eq:Tp_gen}
\end{equation}
where $g_{\alpha,0}$ ($g_{\alpha,1}$) denote the degeneracies of
the ground (first excited) states in sector $\alpha\in\{a,b\}$, respectively.
The quantity $\epsilon_{p}=\varepsilon_{a,0}(x_{p})-\varepsilon_{b,0}(x_{p})$
is the energy difference between the ground states, and $\delta_{a,1}(x_{p})$,
$\delta_{b,1}(x_{p})$ are the excitation gaps within each sector.

If excited states are negligible, this reduces to
\begin{equation}
T_{p}=\frac{\epsilon_{p}}{\ln(g_{a,0}/g_{b,0})},\label{eq:Tp}
\end{equation}
a widely used expression in the literature\citep{pseudo}.
A positive $T_{p}$ occurs only when energetic and entropic contributions
compete: $\epsilon_{p}<0$ with $g_{a,0}<g_{b,0}$, or $\epsilon_{p}>0$
with $g_{a,0}>g_{b,0}$. Otherwise, one sector dominates for all temperatures
and no pseudotransition appears in the canonical ensemble.

The degeneracies determine where the condition $w_{a}=w_{b}$
is met, since the entropic terms $-T\ln g_{\alpha,0}$ shift the effective
free energies of the two sectors. A sector with larger ground-state
multiplicity may therefore dominate even when its energy is higher,
as explored in Sec. \ref{sec:4}. When the degeneracies are equal
this entropic contribution cancels, and the balance condition is determined
by the structure of the low-lying excitations and $T_{p}$ is obtained
from \eqref{eq:Tp_gen}, as discussed in Sec. \ref{sec:5}. Equal
degeneracies do not guarantee a pseudotransition; such behavior occurs
only when the excitation gaps differ sufficiently to produce an exponentially
sharp, yet analytic, crossover.

\subsection{Zero temperature limit along the pseudo-critical
line}

We consider the low-temperature limit restricted
to the pseudo-critical line $T=T_{p}(x)$, defined in the previous
subsection by the condition $w_{a}(x,T_{p})=w_{b}(x,T_{p})$. The
diagonal weights consist of an energetic and an entropic factor: $\varepsilon_{\alpha,0}(x)$
is the ground-state energy of sector $\alpha$, while $g_{\alpha,0}$
counts the number of microstates compatible with that energy. Using
their low-temperature form, the balance condition becomes
\begin{equation}
w_{a}=w_{b}\;\Rightarrow\;g_{a,0}e^{-(\bar{\varepsilon}+\epsilon/2)/T_{p}}=g_{b,0}e^{-(\bar{\varepsilon}-\epsilon/2)/T_{p}},
\end{equation}
where $\bar{\varepsilon}=\frac{\varepsilon_{a,0}(x)+\varepsilon_{b,0}(x)}{2}$
is the average ground-state energy and $\epsilon=\varepsilon_{a,0}(x)-\varepsilon_{b,0}(x)$
is their difference.

Rearranging yields the constrained relation $\frac{g_{a,0}}{g_{b,0}}=e^{\epsilon/T_{p}}$,
which allows us to express the average weight as
\begin{alignat}{1}
\bar{w}= & \frac{1}{2}\left[g_{a,0}e^{-\epsilon/2T_{p}}+g_{b,0}e^{\epsilon/2T_{p}}\right]e^{-\bar{\varepsilon}/T_{p}}\nonumber \\
= & \sqrt{g_{a,0}g_{b,0}}e^{-\bar{\varepsilon}/T_{p}}.
\end{alignat}
The entropy per site associated with this balanced configuration is
therefore
\begin{equation}
\mathcal{S}_{0}^{\mathrm{c}}=\ln\left(\sqrt{g_{a,0}g_{b,0}}\right),\label{eq:S_0^p}
\end{equation}
which is the geometric mean of the ground-state degeneracies.
An alternative rigorous demonstration of $\mathcal{S}_{0}^{c}$ is
given in Appendix \ref{sec:Appdx-B}. 

The entropy \eqref{eq:S_0^p} is valid when the low-temperature
behavior is controlled by two discrete sectors with finite excitation
gaps, so that their asymptotic weights remain accurate along the \textit{pseudo-critical
line}. In this regime the off-diagonal mixing is
exponentially small, and the condition $w_{a}(T_{p})=w_{b}(T_{p})$
determines the leading contribution to the dominant eigenvalue. The
limit $T\to0$ must be taken along this line; at fixed parameters
one sector always dominates and the entropy reduces to $\ln\left(g_{\alpha,0}\right)$.
If additional sectors compete, or if the mixing becomes comparable
to the diagonal weights, the two-sector description fails and Eq.
\eqref{eq:S_0^p} no longer applies.

The residual entropy assumes an intermediate value
because irreducibility ($w_{0}>0$) enforces strictly positive statistical
weights for both sectors along the pseudo-critical line. When the
limit $T\to0$ is taken with $w_{0}>0$, the stationary measure retains
contributions from both sectors and yields $\mathcal{S}_{0}^{c}=\ln\sqrt{g_{a,0}g_{b,0}}$.
In contrast, imposing $w_{0}=0$ before the low-temperature limit
makes the transfer matrix reducible, causing the system to collapse
onto a single dominant sector and producing $\max(\ln g_{a,0},\ln g_{b,0})$.
The two limits do not commute, and the maximal value arises only in
the strictly reducible case.

\subsubsection{Residual entropy inequality}\label{subsec:4-c}

Assuming $g_{a,0}\geqslant g_{b,0}$, without loss of generality,
we define the residual entropies of the adjacent pure sectors as $\mathcal{S}_{0}^{a}=\ln(g_{a,0})$,
$\mathcal{S}_{0}^{\text{b}}=\ln(g_{b,0})$. The interface residual
entropy derived in Eq. \eqref{eq:S_0^p}, results in $\mathcal{S}_{0}^{{\rm c}}=\frac{1}{2}(\mathcal{S}_{0}^{{\rm a}}+\mathcal{S}_{0}^{{\rm b}})$,
then satisfies the inequality

\begin{equation}
\mathcal{S}_{0}^{\text{b}}\leqslant\mathcal{S}_{0}^{\mathrm{c}}\leqslant\mathcal{S}_{0}^{\text{a}},\label{eq:entropy-cond}
\end{equation}
which expresses that the crossover entropy lies between those of the
competing sectors. This is a direct consequence of the nearly block-diagonal
matrix structure and the balancing condition at the pseudotransition.
Violating this inequality precludes a pseudotransition, although satisfying
it alone is not sufficient. This provides a relaxed version of the
criterion previously discussed in Ref.\,\citep{ph-bd}.

The entropy \eqref{eq:entropy-cond} inequality offers a useful diagnostic
tool: it characterizes pseudotransition-like behavior in systems with
finite-range interactions, while preserving analyticity of the free
energy. It emerges naturally from the weak coupling between dominant
irreducible sectors in a nearly block-diagonal transfer matrix.

The interface residual entropy $\mathcal{S}_{0}^{\mathrm{c}}$ measures
the configurational complexity at the crossover. When $\mathcal{S}_{0}^{\mathrm{c}}\in[\mathcal{S}_{0}^{b},\mathcal{S}_{0}^{a}]$,
the sectors contribute comparably, allowing a pseudotransition. In
contrast, models with interface frustration, or symmetry mixing often
yield $\mathcal{S}_{0}^{\mathrm{c}}>\max(\mathcal{S}_{0}^{a},\mathcal{S}_{0}^{b})$,
signaling excess disorder at the interface and forbidding a pseudotransition.

\section{Doniach one-dimensional model}\label{sec:4}

To illustrate the proposed formalism, we consider a representative
model that captures the essential features of pseudotransitions in
one-dimensional systems. Specifically, we study an Ising chain with
internal degeneracy, where one spin state carries a macroscopic multiplicity.
In particular, it shares structural features with the model proposed
by Doniach\citep{Doniach} to describe lipid chain ordering in biomembranes\citep{Mosgaard}.
While our focus here is on the spectral and thermodynamic aspects
of degeneracy-induced pseudotransitions, this biological motivation
reinforces the broader relevance of such models.  Despite its simplicity,
this model is relevant to physical systems such as biomembranes, hydrogen-bonded
chains, and DNA denaturation (for detail see Section 4.2 of Ref.\,\citep{tiago}),
and it exhibits pseudo-critical behavior closely related to that seen
in Potts-like and Zimm-Bragg-type models \citep{panov2021,badasyan}.
Furthermore, several more elaborated models can be mapped to similar
effective forms via decoration transformations \citep{Dec-trnsf},
making this a useful minimal prototype for analyzing pseudotransitions.

\subsection{Hamiltonian and transfer matrix}

In Doniach’s original model\,\citep{Doniach}, the system is a one-dimensional
chain of variables $s_{i}=\pm1$, representing two conformational
states of a lipid chain e.g., $s_{i}=+1$ for the ordered ($a$) state
and $s_{i}=-1$ for the disordered ($b$) state. The simplified Hamiltonian
is
\begin{equation}
\mathcal{H}=-J\sum_{i}s_{i}s_{i+1}-h_{\text{eff}}(T)\sum_{i}s_{i},
\end{equation}
where $J>0$ favors alignment (e.g., adjacent $a$-states), and $h_{\text{eff}}(T)$
is a temperature-dependent effective field that represents entropic
and energetic contributions. Specifically, $h_{\text{eff}}(T)=\frac{1}{2}(E_{s}+\pi\Delta A+zW-T\mathcal{S}_{\text{eff}}),$
with $E_{s}$ the excitation energy from state $a$ to $b$, $\pi\Delta A$
the mechanical work due to area expansion, $zW$ the interchain interaction
energy, and $\mathcal{S}_{\text{eff}}=k_{B}\ln\left(m_{\text{eff}}\right)$
the configurational entropy of the $b$-state. 

Now we can interpret as internal degeneracy, the state $s_{i}=-1$
has multiplicity $m_{\text{eff}}\geqslant1$, while $s_{i}=+1$ remains
non-degenerate. Although this modification does not alter the energy
spectrum, it modifies the statistical weights in the partition function
by introducing a degeneracy factor associated with each occurrence
of $s_{i}=-1$.

An effective way to represent this asymmetry is by redefining the
Hamiltonian to include an entropic term:

\begin{equation}
\mathcal{H}_{{\rm eff}}=-J\sum_{i=1}^{N}s_{i}s_{i+1}-\sum_{i=1}^{N}\left[hs_{i}+(1-s_{i})\frac{\ln\left(m_{\text{eff}}\right)}{2\beta}\right].
\end{equation}
where $\beta=1/T$ and Boltzmann’s constant is set to unity. This
form shifts the energy associated with the $s_{i}=1$ state by a temperature-dependent
term, effectively encoding the degeneracy $m_{\text{eff}}$ through
a modified Boltzmann factor.

The resulting $2\times2$ transfer matrix becomes

\begin{equation}
\mathbf{V}=\begin{pmatrix}e^{\beta(J+h)} & m_{\text{eff}}\,e^{-\beta(J-h)}\\
e^{-\beta(J+h)} & m_{\text{eff}}\,e^{\beta(J-h)}
\end{pmatrix},\label{eq:Ising-matx}
\end{equation}
where the matrix elements correspond to the Boltzmann weights of neighboring
spin pairs $(s_{i},s_{i+1})$. The degeneracy $m_{\text{eff}}$ modifies
the statistical weight of configurations involving the $-1$ state,
introducing entropic competition between sectors.

This simple modification captures key features relevant to pseudotransition
behavior and allows full analytical treatment, making it a suitable
minimal model for illustrating the general framework.

According to Sec.\,\ref{subsec:2-a}, we can interpret the corresponding
block-diagonal matrix, which in this case consists of the diagonal
elements of Eq.\,\eqref{eq:Ising-matx}, with associated dominant
configurations $|\psi_{a}\rangle=|+\rangle$ and $|\psi_{b}\rangle=|-\rangle$,
and corresponding weights $w_{a}=e^{\beta(J+h)}$ and $w_{b}=m_{\text{eff}}\,e^{\beta(J-h)}$,
respectively. Evidently, the off-diagonal elements of matrix $\mathbf{V}$
become negligible in the limit $\beta\to\infty$ and $J>0$. In this
regime, the system undergoes a zero-temperature phase transition between
the fully spin-down state $|\psi_{b}\rangle$ and the fully spin-up
state $|\psi_{a}\rangle$. Note that spin-inversion symmetry holds
only when $m_{\text{eff}}=1$.

\subsection{Pseudo-critical temperature and interface residual entropy}

To analyze the thermodynamic behavior near the pseudotransition, we
use the framework developed in Sec.\ref{subsec:3-a}. Identifying
the dominant weights as $w_{a}=e^{\beta(J+h)}$, $w_{b}=m_{\text{eff}}\,e^{\beta(J-h)}$,
and $w_{0}=\sqrt{m_{\text{eff}}}\,e^{-\beta J}$, the spectral structure
is governed by 
\begin{equation}
\bar{w}=\frac{e^{\beta J}}{2}\left(e^{\beta h}+m_{\text{eff}}\,e^{-\beta h}\right),\enspace\sigma=e^{\beta J}\left(e^{\beta h}-m_{\text{eff}}\,e^{-\beta h}\right),
\end{equation}
 and the spectral gap \eqref{eq:zeta-def}, becomes

\begin{equation}
\zeta=\frac{\sqrt{\sigma^{2}+4m_{\text{eff}}\,e^{-2\beta J}}}{\bar{w}}.
\end{equation}
Recall from Sec.\ref{subsec:3-a} that $\zeta$ also controls the
thermodynamic signatures, and small $\zeta$ enhances the pseudotransition.

To determine the pseudo-critical temperature, we apply the condition
$\partial_{h}\zeta=0$, as given by Eq.\eqref{eq:gen-xp}. This condition
reduces to $w_{a}=w_{b}$, which leads to the closed-form expression:

\begin{equation}
T_{p}=\frac{2h_{p}}{\ln(m_{\text{eff}})},\label{eq:T_p}
\end{equation}
consistent with previous results\citep{tiago}. 

We may also compute the value $T_{p^{*}}$ that minimizes $\zeta$
with respect to temperature, i.e., using $\zeta_{T_{p^{*}}}=0$ from
Eq.\eqref{eq:gen-Tp*}. The result reads
\begin{equation}
\left(m_{\text{eff}}-{\rm e}^{2\beta h}\right)h{\rm e}^{4\beta J}+\left(2J+h\right){\rm e}^{2\beta h}+\left(2J-h\right)m_{\text{eff}}=0,
\end{equation}
which, to leading order in the low-temperature regime, becomes:
\begin{alignat}{1}
{\rm e}^{2\beta h_{p}}= & m_{\text{eff}}\,\frac{{\rm e}^{4\beta J}-1+2J/h_{p}}{{\rm e}^{4\beta J}-1-2J/h_{p}},\label{eq:T_p*-cond}\\
{\rm e}^{2\beta h_{p}}\approx & m_{\text{eff}}+\frac{4J\omega}{h_{p}}{\rm e}^{-4\beta J}.\label{eq:T_p-aprx}
\end{alignat}
yielding the approximate correction, in analogy a \eqref{eq:T_p},
we have
\begin{equation}
T_{p^{*}}\approx\frac{2h_{p}}{\ln(m_{\text{eff}})+\frac{4J}{h_{p}}{\rm e}^{-4J/T_{p^{*}}}}.\label{eq:T_p*}
\end{equation}
Since $J>0$ and $\beta\gg1$, the correction term in Eq. \eqref{eq:T_p-aprx}
becomes negligible, and $T_{p}$ and $T_{p^{*}}$ are of the same
leading order.

\begin{figure}
\includegraphics[scale=0.6]{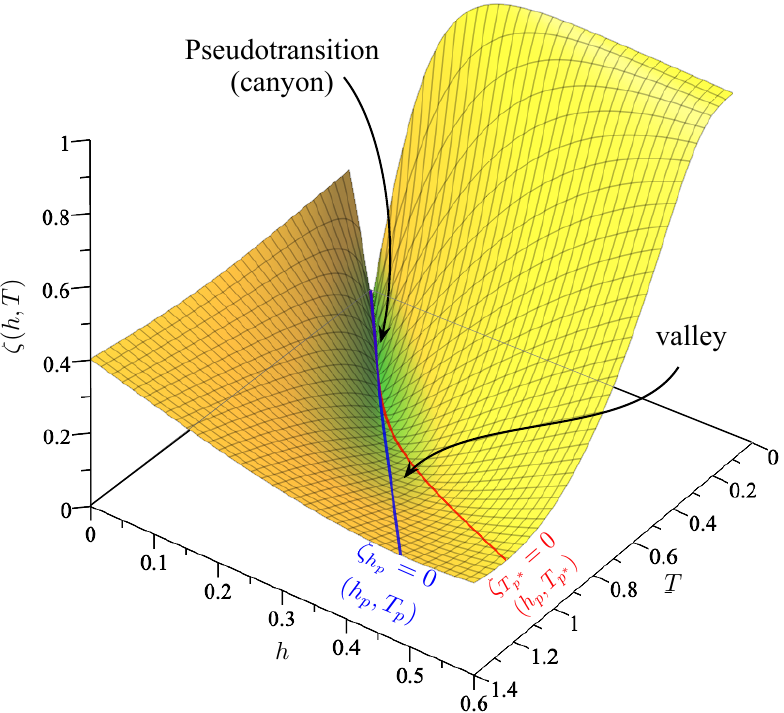}\caption{Surface plot of $\zeta(h,T)$ in the $h-T$ plane for $m_{\text{eff}}=2$
and $J=1$. The blue curve corresponds to the condition $\zeta_{h}=0$
{[}\eqref{eq:T_p}{]}, while the red curve follows $\zeta_{T}=0$
{[}\eqref{eq:T_p*-cond}{]}.}\label{fig:Surface-of-zeta}
\end{figure}

Fig. \ref{fig:Surface-of-zeta} shows the surface $\zeta(h,T)$ for
fixed parameters $m_{\text{eff}}=2$ and $J=1$. The blue curve corresponds
to $\zeta_{h}=0$ {[}see \eqref{eq:T_p}{]} and the red curve to $\zeta_{T}=0$
{[}see \eqref{eq:T_p*-cond}{]}. At higher temperatures, these curves
are well separated, but as temperature decreases, they converge near
a characteristic field $h_{p}$. This convergence defines a narrowing
valley in the ($h,T$)-plane, resembling a “spectral canyon”, with
a floor given by the minimal values of $\zeta$. As $T\to0$, the
gap region becomes increasingly localized, signaling the sharpening
of the pseudotransition. Although the free energy remains analytic
throughout, this spectral structure reflects strong competition between
sectors and underpins the rapid thermodynamic crossover characteristic
of pseudotransitions.

At this point, the Boltzmann weights of the competing spin sectors
become equal. For this model, the corresponding residual entropy is
${\cal S}_{0}^{{\rm c}}=\frac{1}{2}\ln(m_{\text{eff}})=\frac{1}{2}\mathcal{S}_{\text{eff}}$,
which follows directly from \eqref{eq:S_0^p}, in agreement with the
result reported in Ref.\,\citep{tiago}. This residual entropy satisfies
the inequality given in Eq. \eqref{eq:entropy-cond}, 
\begin{equation}
0<{\cal S}_{0}^{{\rm c}}=\frac{1}{2}\mathcal{S}_{\text{eff}}<\mathcal{S}_{\text{eff}},
\end{equation}
consistent with the general criterion discussed in Sec.\ref{subsec:4-c}.
It reflects the partial contribution of degenerate configurations
at the crossover and confirms the presence of a pseudotransition,
characterized by near-degenerate dominant eigenvalues and a nearly
block-diagonal transfer matrix structure.

Surely, this approach can also be applied to higher spins, as explored
in Ref.\,\citep{tiago} for spin-1 systems, which involve a cubic-root
equation.

\section{Mixed spin-1/2 and spin-1 hexagonal nanowire system}\label{sec:5}

The next model we consider contrasts with the previous one. Here,
our goal is to demonstrate how a high-dimensional transfer matrix
can be systematically reduced, first using symmetries, and then through
the nearly block-diagonal structure central to pseudotransition behavior.

\begin{figure}
\includegraphics[scale=0.5]{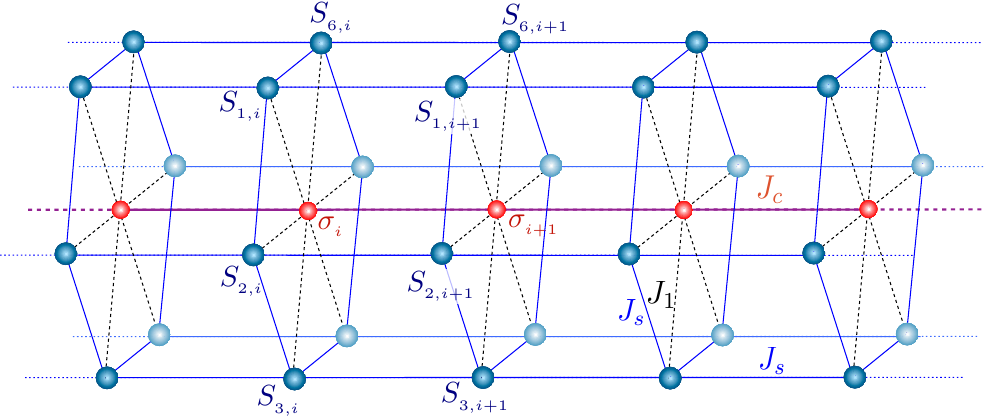}

\caption{Schematic representation of mixed spin-$1/2$ (red) and spin-1 (blue)
hexagonal nanowire. Shell spins interact via $J_{s}$, core-shell
via $J_{1}$, and core-core via $J_{c}$.}\label{fig:nanowire}

\end{figure}

In this context, motivated by experimental results, several studies
have investigated magnetic nanowire systems with core--shell architectures,
revealing distinctive behaviors in different materials. For example,
single-crystal $\mathrm{Co}{{}_3}\mathrm{O}{{}_4}$ nanowires exhibit
two Néel temperatures (56\,K and 73\,K) due to magnetic proximity
effects between core and shell regions\citep{zhang}; large-scale
Co-Zn-P nanowire arrays synthesized via electroless deposition show
characteristic magnetic responses linked to their composition and
structure\citep{yuan}; and Co-Cu nanowire arrays created using galvanic
displacement deposition demonstrate tunable microstructures and magnetic
properties\citep{yang}. Several related theoretical investigations
have been carried out, such as the one developed in Ref.\, \citep{Masrour,Ertas}.
Therefore, we analyze the mixed-spin hexagonal nanowire model studied
in Refs.\,\citep{Rodrigo,Mendes}, consisting of an Ising spin-1/2
(red circles) coupled to spin-1 Blume-Capel variables (blue circles),
as illustrated in Fig.\ref{fig:nanowire}. 

The system is defined on a cylindrical geometry and described by the
Hamiltonian $\mathcal{H}=\sum_{i=1}^{N}H_{i,i+1}$, with 
\begin{alignat}{1}
H_{i,i+1}= & -\sum\limits_{j=1}^{6}\left[J_{s}\left(S_{j,i}S_{j,i+1}+S_{j,i}S_{j+1,i}\right)\right.\nonumber \\
 & \left.J_{1}S_{j,i}\sigma_{i}+J_{c}\sigma_{i}\sigma_{i+1}+DS_{j,i}^{2}\right],\label{eq:H}
\end{alignat}
where $\sigma_{i}\in\left\{ -\tfrac{1}{2},\tfrac{1}{2}\right\} $
denotes the spin-1/2 core spin and $S_{j,i}\in\{-1,0,1\}$ the spin-1
shell spin at position $j$ in the $i$-th hexagonal unit. We assume
periodic boundary conditions. The coupling constants are: $J_{1}$
between Ising and Blume-Capel spins, $J_{c}$ between core Ising spins,
and $J_{s}$ between neighboring shell spins; $D$ is the single-ion
anisotropy (crystal field).

\subsubsection{Ferromagnetic (FM) phase }

For $D>-2.5$, the ground state of the system is ferromagnetic (FM),
with both core and shell spins fully aligned. The corresponding ground-state
energy per unit cell is
\begin{alignat}{1}
E_{FM}= & -\frac{61}{4}-6D.
\end{alignat}
Each unit cell configuration is denoted by
\[
|\mathfrak{a}_{\tau}^{(0)}\rangle=|\tau,\tau,\tau,\tau,\tau,\tau\rangle\otimes|\tfrac{\tau}{2}\rangle,
\]
where $\tau=\pm1$. The first ket corresponds to the shell spin-1
states, while the second ket denotes the core spin-1/2 state. The
full ground state is the symmetric superposition over all $N$ unit
cells:
\begin{equation}
|FM\rangle=2^{-N/2}\bigotimes_{i=1}^{N}\left(|\mathfrak{a}_{+}^{(0)}\rangle_{i}+|\mathfrak{a}_{-}^{(0)}\rangle_{i}\right).
\end{equation}
This state is doubly degenerate due to global spin-inversion symmetry.
Although each configuration is magnetically ordered, the overall thermal
average of the magnetization vanishes: $m_{c}=m_{s}=0$. 

\subsubsection{Core-ferromagnetic (CFM) phase}

For $D<-2.5$, the ground state shifts to a core-ferromagnetic (CFM)
configuration, where the core spins are aligned while the shell spins
are fully suppressed into the $S=0$ state. The corresponding ground-state
energy per unit cell is
\begin{equation}
E_{CFM}=-\frac{1}{4}.
\end{equation}
Each unit cell state is given by
\[
|\mathfrak{b}_{\tau}^{(0)}\rangle,=|0,0,0,0,0\rangle\otimes|\tfrac{\tau}{2}\rangle,
\]
where $\tau=\pm1$. The full ground state takes the form

\begin{equation}
|CFM\rangle=2^{-N/2}\bigotimes_{i=1}^{N}\left(|\mathfrak{b}_{+}^{(0)}\rangle_{i}+|\mathfrak{b}_{-}^{(0)}\rangle_{i}\right),
\end{equation}
This state is also doubly degenerate due to global spin inversion.
Since the shell spins are in non-magnetic $S=0$ states, we have $m_{s}=0$,
and the net core-spin magnetization averages to zero: $m_{c}=0$.

Before deriving the effective thermodynamics, we
summarize the structure of the transfer matrix. The full $1458\times1458$
matrix decomposes into invariant subspaces fixed by the $C_{6v}$
symmetry of the hexagonal shell and by the allowed configurations
of the core spin, as detailed in Sec. \ref{subsec:Reducible-block-transfer}.
This yields a finite set of symmetry-related blocks, the largest being
$92\times92$ as detailed in Ref. \citep{Rodrigo}. Near the pseudo-critical
region, a single block dominates and becomes effectively two-dimensional
due to the competition between two low-energy sectors as detailed
in Sec.\ref{subsec:Nearly-block-diagonal-matrix}. The resulting $2\times2$
matrix reproduces the leading eigenvalues shown in Fig. \ref{fig:leading-eig-app},
confirming the reduction.

\subsection{Reducible block transfer matrix}\label{subsec:Reducible-block-transfer}

We now apply the transfer matrix method to compute the free energy
of the nanowire model defined by the Hamiltonian in Eq.\,\eqref{eq:H}.
The full transfer matrix $\mathbf{V}$ has dimension $1458\times1458$,
with elements defined as
\begin{equation}
\langle\sigma,\{S_{i}\}|\mathbf{V}|\{S'_{i}\},\sigma'\rangle={\rm e}^{-H_{i,i+1}/k_{B}T},\label{transfer}
\end{equation}
where $\{S_{i}\}=\{S_{1},S_{2},S_{3},S_{4},S_{5},S_{6}\}$ denotes
the spin-1 shell configuration of the hexagon at site $i$, and $\sigma\in\left\{ -\tfrac{1}{2},\tfrac{1}{2}\right\} $
is the core spin.

This transfer matrix is reducible due to the geometric and spin symmetries
of the system. In particular, the hexagonal symmetry group $\mathcal{C}_{6}$
acts by cyclic permutation:
\begin{alignat}{1}
\mathcal{C}_{6}|S_{1},S_{2},S_{3},S_{4},S_{5},S_{6}\rangle= & |S_{2},S_{3},S_{4},S_{5},S_{6},S_{1}\rangle,
\end{alignat}
with the full group given by $\mathcal{C}_{6}\mapsto\{I,\mathcal{C}_{6},\mathcal{C}_{6}^{2},\dots,\mathcal{C}_{6}^{5}\}$.
In addition, the system is invariant under spatial reflection $\mathcal{R}$
through the center of the hexagon: 
\begin{alignat}{1}
\mathcal{R}|S_{1},S_{2},S_{3},S_{4},S_{5},S_{6}\rangle= & |S_{6},S_{5},S_{4},S_{3},S_{2},S_{1}\rangle,
\end{alignat}
with $\mathcal{R}\mapsto\{I,\mathcal{R}\}$.

Taking into account the full $C_{6v}$ symmetry, the transfer matrix
decomposes into a direct sum of irreducible subspaces. The shell space
of dimension $3^{6}=729$ can be block-diagonalized under these symmetries,
and the full transfer matrix $\mathbf{V}$ of dimension $1458=2\times729$
can be reduced accordingly.

This symmetry reduction leads to the decomposition:
\begin{equation}
3^{6}\otimes2=184\oplus178\oplus172\oplus3(\oplus160)\oplus6(\oplus74),
\end{equation}
where $n(\oplus d)$ denotes $n$ blocks of dimension $d$. Moreover,
in the absence of external fields, the model is symmetric under global
spin inversion ($\mathbb{Z}_{2}$), $(S_{j}^{z},\sigma^{z})\mapsto(-S_{j}^{z},-\sigma^{z})$,
which allows further reduction of $\mathbf{V}$ into even and odd
parity sectors of dimension $729\times729$. 

Each of these sectors further decomposes as:
\begin{alignat}{1}
3^{6}= & 92\oplus89\oplus86\oplus3(\oplus80)\oplus6(\oplus37)\,.
\end{alignat}
The dominant eigenvalue, which governs the free energy in the thermodynamic
limit, lies in the largest irreducible block of size $92\times92$,
associated with the fully symmetric subspace under rotations, reflections,
and global spin inversion. As shown in Ref.\,\citep{Rodrigo}, this
eigenvalue is non-degenerate and satisfies the Perron-Frobenius theorem,
ensuring analyticity and excluding a true phase transition.

\subsection{Nearly block-diagonal matrix}\label{subsec:Nearly-block-diagonal-matrix}

The mean-field approximation (MFA) \citep{Mendes} indicates a first-order
phase transition at finite temperature near $D=-2.5$, for coupling
constants $J_{1}=J_{s}=J_{c}=1$. However, a subsequent analysis using
the numerical transfer matrix method \citep{Rodrigo} revealed that
the system exhibits only a pseudotransition, with no genuine thermodynamic
singularity.

\subsubsection{0th-order correction}

At zeroth order, we neglect all low-lying excited states and retain
only the dominant ground-state configurations $\{|a^{(0)}\rangle,|b^{(0)}\rangle\}$,
corresponding to the FM and CFM sectors, respectively. In this approximation,
the transfer matrix reduces to a $2\times2$ effective form:
\begin{equation}
\mathbf{V}_{t}^{(0)}=\begin{pmatrix}x^{12}\left(y^{61}\pm y^{11}\right) & y^{17}x^{6}\left(y^{2}\pm1\right)\\
y^{17}x^{6}\left(y^{2}\pm1\right) & \frac{y^{2}\pm1}{y}
\end{pmatrix},\label{eq:V_eff0}
\end{equation}
with $y={\rm e}^{J_{c}/4T}$ and $x={\rm e}^{D/2T}$. The $(+)$ sign
corresponds to the symmetric sector, and the $(-)$ sign to the antisymmetric
one.

As discussed in Sec. \ref{subsec:3-a}, $\zeta=\ln(\lambda_{1}^{(0)}/\lambda_{2}^{(0)})$
reaches a minimum at the pseudo-critical point, providing a local
diagnostic for anomalous thermodynamics.

\begin{figure}
\includegraphics[scale=0.55]{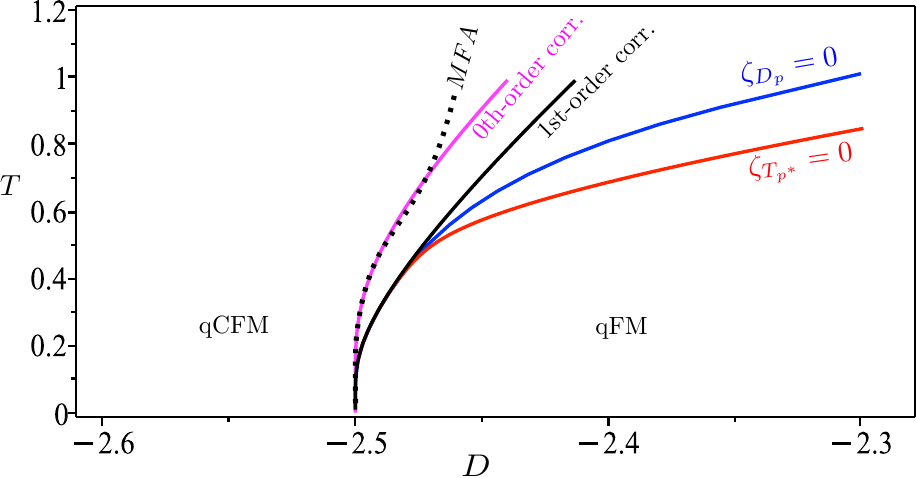}

\caption{Pseudo-critical temperature $T_{p}$ as a function of $D$, assuming
zero magnetic field and fixed $J_{1}=J_{s}=J_{c}=1$. The curves correspond
to different pseudotransition conditions: \eqref{eq:gen-Tp*} (blue
curve), \eqref{eq:gen-xp} (red curve), the 0th-roder correction \eqref{eq:0th-cor-cond}
(pink curve), and 1st-order correction \eqref{eq:1st-corr-con} (black
curve). The dashed line shows the MFA result from \citep{Mendes}.}\label{fig:Pseudo-critical-temperature}
\end{figure}

Comparing matrix \eqref{eq:V_eff0} with the general effective matrix
form given in Eq.\,\eqref{eq:V-eff}, and applying the pseudo-critical
condition \eqref{eq:w1-w2} $w_{a}=w_{b}$, we obtain the following
transcendental equation from the diagonal elements of the symmetric
block:
\begin{equation}
{\rm e}^{\frac{2D_{p}+J_{c}}{4T_{p}}}\left({\rm e}^{\frac{25J_{c}}{2T_{p}}}+1\right)=\left({\rm e}^{\frac{J_{c}}{2T_{p}}}+1\right).\label{eq:0th-cor-cond}
\end{equation}
Solving this equation numerically for fixed $J_{c}=1$, we obtain
the pseudo-critical temperature $T_{p}(D)$, plotted as the pink curve
in Fig.\ref{fig:Pseudo-critical-temperature}. Surprisingly, this
analytical result matches well with the MFA prediction (dashed line)
from Ref.\,\citep{Mendes}, which was originally interpreted as signaling
a first-order phase transition.

\subsubsection{1st-order correction}\label{subsec:1st-order-correction}

In the low-temperature regime, the system is primarily governed by
ground-state configurations. However, as temperature increases, thermal
fluctuations begin to populate low-lying excited states. For simplicity,
we label the dominant sectors as regions $a$ and $b$, following
the conventions introduced in Sec.\,\ref{subsec:2-a}.

The first excited states within each region can be defined using the
following symmetric combinations. For region $a$, we define:
\begin{alignat}{1}
|\mathfrak{a}_{\tau}^{(1)}\rangle= & \tfrac{1}{\sqrt{6}}\bigl(|\tau\tau\tau\tau\tau0\rangle+\sqrt{5}|\tau\tau\tau0\tau0\rangle\bigr)\otimes|\tfrac{\tau}{2}\rangle,
\end{alignat}
which represents the leading fluctuation above the FM ground state.
Similarly, for region $b$, we denote
\begin{alignat}{1}
|\mathfrak{b}_{\tau}^{(1)}\rangle= & \tfrac{1}{\sqrt{6}}(|00000\tau\rangle+\sqrt{5}|000\tau0\tau\rangle)\otimes|\tfrac{\tau}{2}\rangle,
\end{alignat}
corresponding to the dominant fluctuation above the CFM ground state.

To construct the truncated transfer matrix, we apply spin-inversion
symmetry to build the parity-even combinations
\begin{alignat}{1}
|a^{(k)}\rangle= & \tfrac{1}{\sqrt{2}}\left(|\mathfrak{a}_{+}^{(k)}\rangle+|\mathfrak{a}_{-}^{(k)}\rangle\right),\label{eq:a_(k)}\\
|b^{(k)}\rangle= & \tfrac{1}{\sqrt{2}}\left(|\mathfrak{b}_{+}^{(k)}\rangle+|\mathfrak{b}_{-}^{(k)}\rangle\right),\label{eq:b_(k)}
\end{alignat}
with $k=0,1$. Using the truncated basis $\{|a^{(0)}\rangle,|a^{(1)}\rangle,|b^{(0)}\rangle,|b^{(1)}\rangle\}$,
the irreducible transfer matrix becomes 
\begin{equation}
\mathbf{V}_{t}^{(1)}=\begin{pmatrix}v_{1,1} & v_{1,2} & v_{1,3} & v_{1,4}\\
v_{1,2} & v_{2,2} & v_{2,3} & v_{2,4}\\
v_{1,3} & v_{2,3} & v_{3,3} & v_{3,4}\\
v_{1,4} & v_{2,4} & v_{3,4} & v_{4,4}
\end{pmatrix},\label{eq:red-mat}
\end{equation}
with matrix elements given by
\begin{alignat}{1}
v_{1,1}= & x^{12}\left(y^{61}\pm y^{11}\right),\nonumber \\
v_{1,2}= & y^{10}x^{11}\left(y^{42}\pm1\right)\sqrt{6},\nonumber \\
v_{1,3}= & y^{14}x^{7}\left(y^{10}\pm1\right)\sqrt{6},\nonumber \\
v_{1,4}= & y^{17}x^{6}\left(y^{2}\pm1\right),\nonumber \\
v_{2,2}= & x^{10}y^{5}\left(y^{42}5y^{38}\pm5y^{4}\pm1\right),\label{eq:rijs}\\
v_{2,3}= & x^{6}y^{9}\left(5y^{10}+y^{6}\pm y^{4}\pm5\right),\nonumber \\
v_{2,4}= & y^{12}x^{5}\left(y^{2}\pm1\right)\sqrt{6},\nonumber \\
v_{3,3}= & \frac{x^{2}\left(y^{10}+5y^{6}\pm5y^{4}\pm1\right)}{y^{3}},\nonumber \\
v_{3,4}= & x\left(y^{2}\pm1\right)\sqrt{6},\nonumber \\
v_{4,4}= & \frac{y^{2}\pm1}{y},\nonumber 
\end{alignat}
with $y={\rm e}^{J_{c}/4T}$ and $x={\rm e}^{D/2T}$. The $(+)$ sign
corresponds to the symmetric sector and the $(-)$ sign to the antisymmetric
one.

We now rewrite the truncated transfer matrix $\mathbf{V}_{t}^{(1)}$
in block form, separating regions $a$ and $b$ as:
\begin{equation}
\mathbf{V}_{t}^{(1)}=\begin{pmatrix}\mathbf{A} & \mathbf{C}\\
\mathbf{C}^{\mathrm{T}} & \mathbf{B}
\end{pmatrix},
\end{equation}
where 
\[
\mathbf{A}=\begin{pmatrix}v_{1,1} & v_{1,2}\\
v_{1,2} & v_{2,2}
\end{pmatrix},\:\mathbf{B}=\begin{pmatrix}v_{3,3} & v_{3,4}\\
v_{3,4} & r_{4,4}
\end{pmatrix},\,\mathbf{C}=\begin{pmatrix}v_{1,3} & v_{1,4}\\
v_{2,3} & v_{2,4}
\end{pmatrix}.
\]
If off-diagonal block $\mathbf{C}$ is sufficiently small in the regime
of interest, we can neglect its contribution and approximate $\mathbf{V}_{t}^{(1)}$
as block-diagonal. In this limit, the eigenvalues and eigenstates
can be obtained independently for each sector.

\paragraph{Sector $a$ (quasi-ferromagnetic):}

Diagonalizing $\mathbf{A}$, the symmetric eigenvalue is given by
\begin{equation}
w_{a}=\frac{v_{1,1}+v_{2,2}+\sqrt{(v_{1,1}-v_{2,2})^{2}+4v_{1,2}^{2}}}{2},
\end{equation}
with corresponding eigenstate
\begin{equation}
|\psi_{a}\rangle=\sin(\theta_{a})|a^{(0)}\rangle+\cos(\theta_{a})|a^{(1)}\rangle,
\end{equation}
where $\tan(2\theta_{a})=\frac{2v_{1,2}}{v_{1,1}-v_{2,2}}.$ The state
$|\psi_{a}\rangle$ is a coherent mixture of the four configurations
$\{|\mathfrak{a}_{\pm}^{(0)}\rangle,|\mathfrak{a}_{\pm}^{(1)}\rangle\}$,
and defines the quasi-ferromagnetic (qFM) collective state:
\begin{equation}
|qFM\rangle=\bigotimes_{i=1}^{N}|\psi_{a}\rangle_{i}.
\end{equation}
At low temperatures, $|\psi_{a}\rangle$ is dominated by the ground
state $|\mathfrak{a}\tau^{(0)}\rangle$, with small thermal admixture
from $|\mathfrak{a}\tau^{(1)}\rangle$. Since $\theta_{a}$ depends
on temperature, the qFM state evolves with thermal fluctuations.

\paragraph{Sector $b$ (quasi-core-ferromagnetic):}

Similarly, diagonalizing $\mathbf{B}$, we obtain the symmetric eigenvalue
\begin{equation}
w_{b}=\frac{v_{3,3}+v_{4,4}+\sqrt{(v_{3,3}-v_{4,4})^{2}+4v_{3,4}^{2}}}{2},\label{eq:wb-nano-apx}
\end{equation}
with eigenstate
\begin{equation}
|\psi_{b}\rangle=\sin(\theta_{b})|b^{(0)}\rangle+\cos(\theta_{b})|b^{(1)}\rangle,\label{eq:wb-ste}
\end{equation}
with $\tan(2\theta_{b})=\frac{2v_{3,4}}{v_{3,3}-v_{4,4}}.$ The state
$|\psi_{b}\rangle$ is a mixture of states $\{|\mathfrak{b}_{\pm}^{(0)}\rangle,|\mathfrak{b}_{\pm}^{(1)}\rangle\}$,
and defines the quasi-core-ferromagnetic (qCFM) state as
\begin{equation}
|qCFM\rangle=\bigotimes_{i=1}^{N}|\psi_{b}\rangle_{i}.
\end{equation}
As with qFM, the qCFM state incorporates temperature-dependent fluctuations
through $\theta_{b}(T)$. Both qFM and qCFM reflect entropically dressed
ground states that become relevant as temperature increases, and together
they define the thermodynamic sectors responsible for the pseudotransition.

\paragraph{Effective transfer matrix:}

Using the truncated matrix approach, we can estimate the largest eigenvalues
of the nanowire model near the anomalous region with good accuracy.
To do so, we construct an effective $2\times2$ irreducible matrix
in the basis $\{|\psi_{a}\rangle,|\psi_{b}\rangle\}$, now including
off-diagonal coupling terms, as introduced in Sec.\,\ref{subsec:2-a}.
The resulting effective transfer matrix reads:

\begin{equation}
\mathcal{V}^{(1)}=\begin{pmatrix}w_{a} & \langle\psi_{a}|\mathbf{C}|\psi_{b}\rangle\\
\langle\psi_{b}|\mathbf{C}^{{\rm T}}|\psi_{a}\rangle & w_{b}
\end{pmatrix}.
\end{equation}
The corresponding dominant eigenvalues of the transfer matrix are

\begin{equation}
\lambda_{1,2}^{(1)}=\tfrac{1}{2}\Bigl(w_{a}+w_{b}\pm\sqrt{(w_{a}-w_{b})^{2}+4w_{0}^{2}}\Bigr),
\end{equation}
where the off-diagonal overlap is defined by $\langle\psi_{a}|\mathbf{C}|\psi_{b}\rangle\langle\psi_{b}|\mathbf{C}^{{\rm T}}|\psi_{a}\rangle=w_{0}^{2}$. 

To determine the pseudo-critical temperature, we impose the condition
$w_{a}=w_{b}$, analogous to Eq.\,\eqref{eq:w1-w2}. Using the explicit
eigenvalue expressions from the symmetric blocks $\mathbf{A}$ and
$\mathbf{B}$, this yields the transcendental equation
\begin{equation}
\frac{v_{1,1}+v_{2,2}+\sqrt{(v_{1,1}-v_{2,2})^{2}+4v_{1,2}^{2}}}{v_{3,3}+v_{4,4}+\sqrt{(v_{3,3}-v_{4,4})^{2}+4v_{3,4}^{2}}}=1.\label{eq:1st-corr-con}
\end{equation}
where all matrix elements $v_{i,j}$ are defined in Eq. \eqref{eq:rijs}
(with the (+) sign). Solving this equation yields the pseudo-critical
temperature $T_{p}$ as a function of $D$. 

The first-order correction result is shown in Fig.\ref{fig:Pseudo-critical-temperature}
as the black solid curve. It differs notably from the zeroth-order
approximation (pink curve) and the MFA prediction, although all three
converge in the low-temperature limit $T\to0$. The same figure also
displays two exact conditions: the blue curve corresponds to $\zeta_{T_{p^{*}}}=0$,
obtained from Eq.\eqref{eq:gen-Tp*}, and the red curve corresponds
to $\zeta_{D_{p}}=0$, derived from Eq. \eqref{eq:gen-xp}.

For temperatures $T\lesssim0.5$, we observe that $T_{p}-T_{p^{*}}\to0$,
and the blue and red curves nearly coincide, forming a deep canyon-like
structure, as seen in Fig.\,\ref{fig:Pseudo-critical-temperature}.
As temperature increases, the curves gradually diverge, outlining
a valley. Notably, our first-order approximation tracks both minimized
curves closely across the entire range, confirming that the first-order
effective matrix yields an accurate estimate of the pseudo-critical
temperature $T_{p}(D)$ throughout the anomalous region.

\begin{figure}

\includegraphics[scale=0.7]{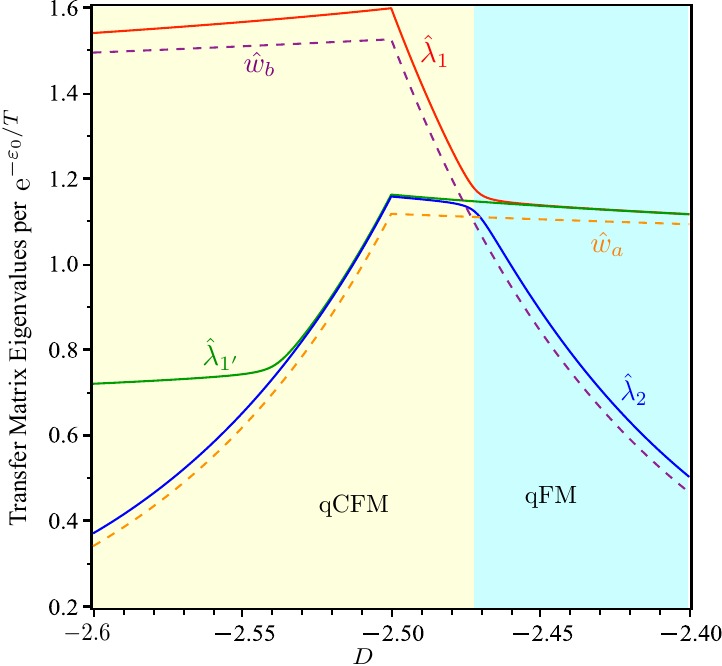}\caption{\textcolor{red}{{} }Leading eigenvalues of the transfer matrix as a
function of $D$ at fixed temperature $T=0.5$. Solid curves represent
exact numerical results: the red curve shows $\hat{\lambda}_{1}$,
the blue curve shows $\hat{\lambda}_{2}$, and the orange curve corresponds
to $\hat{\lambda}_{1'}$. Dashed lines indicate the first-order approximations
$\hat{w}_{a}$ and $\hat{w}_{b}$ from the truncated matrix analysis.}\label{fig:leading-eig-app}
\end{figure}

To assess the accuracy of the truncated effective matrix with first-order
corrections, we compare its predictions to the exact leading eigenvalues
of the full transfer matrix. Fig.\ref{fig:leading-eig-app} shows
the normalized eigenvalues $\hat{\lambda}_{i}=\lambda_{i}\,e^{\varepsilon_{0}/T}$,
where $\varepsilon_{0}=\min(-\tfrac{1}{4},-\tfrac{61}{4}-6D)$ is
the ground-state energy used to regularize the scale. For reference,
we also include the first-order approximations $\hat{w}_{a}=w_{a}\,e^{\varepsilon_{0}/T}$
and $\hat{w}_{b}=w_{b}\,e^{\varepsilon_{0}/T}$ obtained from the
reduced $4\times4$ matrix defined in Sec.\,\ref{subsec:1st-order-correction}.

Fig.\ref{fig:leading-eig-app} illustrates the leading eigenvalues$\hat{\lambda}_{1}$
(solid red curve) and $\hat{\lambda}_{2}$ (solid blue curve), corresponding
to the exact numerical eigenvalues computed from the full $92\times92$
symmetry-reduced transfer matrix. An avoided crossing is observed
near the pseudo-critical value of $D$. The green curve represents
$\hat{\lambda}_{1'}$, the largest eigenvalue of the $92\times92$
antisymmetric sector of the transfer matrix, which corresponds to
the second-largest eigenvalue of the full system. This sector is obtained
through symmetry reduction alone, without any truncation or approximation.

Interestingly, $\hat{\lambda}_{1'}$ remains nearly flat in the vicinity
of the pseudotransition and is largely insensitive to variations in
$D$. Despite being the second-largest eigenvalue in magnitude, its
thermodynamic contribution is suppressed because it originates from
a sector orthogonal to the dominant symmetric eigenstates. This highlights
that $\hat{\lambda}_{2}$, not $\hat{\lambda}_{1'}$, governs the
crossover with $\hat{\lambda}_{1}$, and confirms that the first-order
approximation accurately captures the structure and evolution of the
relevant eigenstates near the anomalous region. For additional detail
see Ref. \citep{Rodrigo}.

\paragraph{Residual entropy inequality:}

Finally, we verify the residual entropy inequality condition. Since
both the FM and CFM phases are non-macroscopically degenerate, we
have $g_{a}=g_{b}=1$, which implies $\mathcal{S}_{a}=\mathcal{S}_{b}=0$.
Consequently, the interface residual entropy \eqref{eq:S_0^p} becomes
$\mathcal{S}_{0}^{\mathrm{c}}=\ln\left(\sqrt{g_{a}g_{b}}\right)=0$,
and the inequality \,\eqref{eq:entropy-cond} is exactly satisfied
\begin{equation}
\mathcal{S}_{0}^{\text{b}}=\mathcal{S}_{0}^{\mathrm{c}}=\mathcal{S}_{0}^{\text{a}}=0.
\end{equation}
There is no entropic competition between sectors, and the interface
behaves thermodynamically like the surrounding phases. In contrast
to models where interface frustration leads to $\mathcal{S}_{0}^{\mathrm{c}}>\max(\mathcal{S}_{0}^{a},\mathcal{S}_{0}^{b}),$
here the residual entropy remains zero throughout.

\section{Conclusion}\label{sec:6}

This work establishes a unifying mechanism for pseudotransitions in
one-dimensional systems: they emerge from irreducible but nearly block-diagonal
transfer matrices. In the low temperature region, off-diagonal elements
coupling distinct symmetry sectors become exponentially suppressed
due to high excitation energies. Although these elements remain nonzero,
ensuring strict irreducibility as required by the Perron-Frobenius
theorem, their small magnitude makes the matrix behave as effectively
reducible over finite temperature ranges. We define this regime as
nearly block-diagonal irreducibility, in which the spectral and thermodynamic
behavior closely resembles that of a reducible system, despite the
absence of genuine singularities.

The key insight is that irreducibility suppresses nonanalyticities,
preventing true phase transitions by forbidding eigenvalue crossings.
Yet, when the dominant eigenvalues from weakly coupled sectors approach
each other, sharp but analytic thermodynamic crossovers arise. This
pseudotransition reflect avoided crossings driven by low-temperature
spectral competition, not actual symmetry breaking. Phase transitions
occur only when irreducibility is lost, i.e., when inter-sector couplings
vanish, allowing disconnected matrix blocks and exact level crossings.

To formalize this, we analyzed the spectral gap function $\zeta(x,T)=\ln(\lambda_{1}/\lambda_{2})$
and derived general pseudo-critical condition based on its extrema.
This condition emerge from eigenvalue competition between leading
symmetry-reduced sectors whose coupling is too weak to induce mixing
but sufficient to lift degeneracies. The mechanism was demonstrated
in two models. In the Ising chain with internal degeneracy (Doniach
model\citep{Doniach}), we derived exact expressions for the pseudo-critical
temperature and residual entropy, revealing how configurational imbalance
drives sharp entropy plateaus. In the mixed spin-1/2 and spin-1 hexagonal
nanowire, symmetry reduction yielded dominant blocks whose interaction
could be described by a small effective matrix. First-order corrections
captured the pseudotransition line with excellent accuracy, even as
full irreducibility was preserved. 

While the emergence of nearly block-diagonal transfer matrices may
seem mathematically expected in low-temperature limits, the thermodynamic
consequences of such structures, particularly the sharp yet analytic
crossovers observed in several theoretical models, have not been previously
identified under a general framework. Our approach clarifies how this
common spectral structure underlies a wide variety of seemingly isolated
or curious behaviors in the literature\citep{pseudo,w-yin-prl,w-yin-prb,w-yin-prr,krokhmalskii2021,panov2021,chapman2024,katarina},
offering a simple and physically intuitive explanation for pseudotransition
phenomena, which in this framework acquire a precise spectral definition
through the minimal gap between dominant transfer-matrix eigenvalues. 

In summary, pseudotransitions originate not from broken analyticity,
but from the interplay of spectral proximity and suppressed coupling
within irreducible transfer matrices. Nearly block-diagonal irreducibility
thus provides a natural explanation for sharp thermodynamic responses
in 1D systems, without contradicting the general absence of phase
transitions\citep{cuesta}. The present analysis is
restricted to discrete energy spectra, and the extension to continuous
spectra will be addressed elsewhere.

\section*{Acknowledgments}
This work was partially supported by the Brazilian agencies CNPq and
FAPEMIG.

\appendix

\section{Derivation of pseudo-critical temperature }\label{subsec:Appendix-A}

This appendix presents the algebraic derivation of
the pseudo-critical conditions using the discrete spectral structure
of a two-sector transfer matrix and the properties of the spectral
gap $\zeta(x,T)$. These assumptions ensure that the low-temperature
asymptotics of the sector weights remain valid.

For each sector we write the effective matrix element as a Boltzmann
sum over internal excitations,
\begin{equation}
w_{\nu}(x,T)=g_{\nu,0}e^{-\varepsilon_{\nu,0}(x)/T}\eta_{\nu}(x,T),\label{eq:w_=00007B=00005Cnu=00007D-appx}
\end{equation}
with $\nu=\{a,0,b\}$, where the correction factor is 
\begin{equation}
\eta_{\nu}(x,T)=1+\sum_{k=1}\hat{g}_{\nu,k}{\rm e}^{-\delta_{\nu,k}(x)/T},
\end{equation}
collects the contributions of excited states. Here $\hat{g}_{\nu,k}=\frac{g_{\nu,k}}{g_{\nu,0}}$
denotes the relative degeneracy of the $k$-th excited level, and
$\delta_{\nu,k}(x)=\varepsilon_{\nu,k}(x)-\varepsilon_{\nu,0}(x)$
being the excitation gap. 

In the low-temperature regime $T\ll\delta_{\nu,1}$,
all excited states are exponentially suppressed and the correction
factor satisfies $\eta_{\nu}(x,T)\approx1$. Under these conditions
the dominant sector weights reduce to their ground-state contributions,
and the pseudo-critical condition can be derived in closed form.

The temperature $T$ and $x$-derivatives of $w_{\nu}$ become
\begin{equation}
\frac{\partial w_{\nu}}{\partial T}=\frac{\varepsilon_{\nu,0}}{T^{2}}w_{\nu},\quad\frac{\partial w_{\nu}}{\partial x}=-\frac{w_{\nu}}{T}\frac{\partial\varepsilon_{\nu,0}}{\partial x}.\label{eq:diff-ws}
\end{equation}
Here we assume the mean and difference of ground-state energies as
$\bar{\varepsilon}=\frac{\varepsilon_{a,0}(x)+\varepsilon_{b,0}(x)}{2}$
and $\epsilon=\varepsilon_{a,0}(x)-\varepsilon_{b,0}(x)$, both of
which depend on $x$. Using the notation in the main text and the
result from \eqref{eq:diff-ws}, we compute
\begin{alignat}{1}
\frac{\partial\sigma}{\partial T}= & \frac{1}{T^{2}}\left(\bar{\varepsilon}\sigma+\epsilon\bar{w}\right),\\
\frac{\partial\bar{w}}{\partial T}= & \frac{1}{T^{2}}\left(\bar{\varepsilon}\bar{w}+\tfrac{1}{4}\epsilon\sigma\right),\\
\frac{\partial w_{0}}{\partial T}= & \frac{1}{T^{2}}\varepsilon_{0,0}w_{0},
\end{alignat}
Substituting into Eq. \eqref{eq:zeta'}, the temperature derivative
of the spectral gap becomes 
\begin{equation}
\frac{\partial\zeta(x,T)}{\partial T}=\frac{\epsilon\sigma\left(w_{a}w_{b}-w_{0}^{2}\right)+4(\bar{\varepsilon}-\varepsilon_{0,0})w_{0}^{2}\bar{w}}{T^{2}\bar{w}^{2}\,\sqrt{4w_{0}^{2}+\sigma^{2}}}.\label{eq:cond-Tp*}
\end{equation}
Assuming $x=x_{p}$ fixed, the anti-crossing (pseudotransition) occurs
when $\zeta(x_{p},T)$ reaches a minimum at temperature $T=T_{p^{*}}$.
Therefore, from \eqref{eq:cond-Tp*} and \eqref{eq:cond-T_p*}, the
pseudotransition condition results in

\begin{equation}
\epsilon\sigma\left(w_{a}w_{b}-w_{0}^{2}\right)+4(\bar{\varepsilon}-\varepsilon_{0,0})w_{0}^{2}\bar{w}=0.\label{eq:gen-Tp*}
\end{equation}

This expression defines the general pseudo-critical condition in the
presence of two competing gapped sectors with weak coupling $w_{0}$.
In the low-temperature limit, $w_{0}\ll w_{a}\sim w_{b}$, and $w_{0}\to0$
as $T\to0$, but remains nonzero for finite $T$. The sign of $\sigma$
may vary and even vanish near the crossover, depending on $T$ and
$x$.

A similar condition arises from the derivative with respect to $x$.
Using:

\begin{alignat}{1}
\frac{\partial\sigma}{\partial x}= & -\frac{1}{T}\left(\bar{\varepsilon}'\sigma+\epsilon'\bar{w}\right),\\
\frac{\partial\bar{w}}{\partial x}= & -\frac{1}{T}\left(\bar{\varepsilon}'\bar{w}+\tfrac{1}{4}\epsilon'\sigma\right),\\
\frac{\partial w_{0}}{\partial x}= & -\frac{1}{T}\varepsilon'_{0,0}w_{0},
\end{alignat}
where the prime denotes derivative with respect to $x$, and $\bar{\varepsilon}'=\frac{\varepsilon'_{a,0}(x)+\varepsilon'_{b,0}(x)}{2}$
and $\epsilon'=\varepsilon'_{a,0}(x)-\varepsilon'_{b,0}(x)$, we obtain,
from eq.\eqref{eq:zeta'}, the derivative of $\zeta(x,T)$ with respect
to $x$ becomes
\begin{alignat}{1}
\frac{\partial\zeta(x,T)}{\partial x}= & \frac{-\epsilon'\sigma\left(w_{a}w_{b}-w_{0}^{2}\right)-4(\bar{\varepsilon}'-\varepsilon'_{0,0})w_{0}^{2}\bar{w}}{T\bar{w}^{2}\,\sqrt{4w_{0}^{2}+\sigma^{2}}}.\label{eq:cond-xp}
\end{alignat}

Thus, the anti-crossing occurs when $\zeta(x,T_{p})$ reaches a minimum
at a specific value $x_{p}$. From \eqref{eq:cond-xp} and \eqref{eq:cond-x_p},
this condition leads to the following expression

\begin{equation}
\epsilon'\sigma\left(w_{a}w_{b}-w_{0}^{2}\right)+4(\bar{\varepsilon}'-\varepsilon'_{0,0})w_{0}^{2}\bar{w}=0.\label{eq:gen-xp}
\end{equation}

In summary, Eq. \eqref{eq:w_=00007B=00005Cnu=00007D-appx}
and the resulting formulas hold in the regime $T\ll\delta_{a,1},\delta_{b,1}$
and $w_{0}\ll w_{a},w_{b}$, that is, within the two-sector low-temperature
window where pseudotransitions arise. The approach assumes a discrete
local spectrum and does not extend to continuous or gapless cases,
for which the hierarchy of Boltzmann weights assumed here no longer
holds.

\section{Entropy at the pseudo-critical point}\label{sec:Appdx-B}

This appendix provides a compact derivation of
the entropy evaluated at the pseudo-critical temperature $T_{p}$.
All quantities and expressions introduced in Sec. \ref{sec:3} (sector
weights, derivatives, eigenvalues $\lambda_{1,2}$, and the low-temperature
forms) are used here.

\subsection{Derivative of the dominant eigenvalue}

Differentiating the largest $\lambda_{1}(T)$ obtained
from \eqref{eq:lamb_12} and then imposing the balance condition $w_{a}(T_{p})=w_{b}(T_{p})$
yields 
\begin{equation}
\lambda'_{1}(T_{p})=\frac{w'_{a}(T_{p})+w'_{b}(T_{p})}{2}+w'_{0}(T_{p}).
\end{equation}
Along the pseudo-critical line, the diagonal weights satisfy the Eq.\eqref{eq:diff-ws}
we have $w_{\nu}'(T)=w_{\nu}(T)\frac{\varepsilon_{\nu,0}}{T^{2}},$so
that
\begin{equation}
\frac{w'_{a}(T_{p})+w'_{b}(T_{p})}{2}=\frac{w(T_{p})}{2T_{p}^{2}}\bigl(\varepsilon_{a,0}+\varepsilon_{b,0}\bigr),
\end{equation}
where $w(T_{p})=w_{a}(T_{p})=w_{b}(T_{p})$. The off-diagonal weight
has the low-temperature form $w_{0}(T)\sim\exp[-\varepsilon_{0,0}/T]$,
hence using \eqref{eq:diff-ws} we have $w'_{0}(T_{p})=w_{0}(T_{p})\frac{\varepsilon_{0,0}}{T_{p}^{2}}$.
The ratio of the off-diagonal derivative to the diagonal one is therefore
\begin{equation}
\frac{w'_{0}(T_{p})}{w'_{a}(T_{p})+w'_{b}(T_{p})}=\frac{w_{0}(T_{p})}{w(T_{p})}\cdot\frac{\varepsilon_{0,0}}{\varepsilon_{a,0}+\varepsilon_{b,0}}.\label{eq:ratios-ws}
\end{equation}
Because pseudo-transition behavior requires $w_{0}(T_{p})\ll w(T_{p})$,
and the energy ratio is $\mathcal{O}(1)$, so the expression in \eqref{eq:ratios-ws}
is exponentially small. Hence 
\begin{equation}
w'_{0}(T_{p})\ll w'_{a}(T_{p})+w'_{b}(T_{p}),
\end{equation}
and to leading order 
\begin{equation}
\lambda'_{1}(T_{p})\simeq\frac{w(T_{p})}{2T_{p}^{2}}\bigl(\varepsilon_{a,0}+\varepsilon_{b,0}\bigr).
\end{equation}

\subsection{Entropy at the pseudo-critical point}

Using
\begin{equation}
\mathcal{S}(T)=\ln\lambda_{1}(T)+T\frac{\lambda'_{1}(T)}{\lambda_{1}(T)},
\end{equation}
and $\lambda_{1}(T_{p})=w(T_{p})+\mathcal{O}(w_{0})$, we obtain
\begin{equation}
\mathcal{S}(T_{p})\simeq\ln w(T_{p})+\frac{\varepsilon_{a,0}+\varepsilon_{b,0}}{2T_{p}}.\label{eq:S-Tp-apx}
\end{equation}
At low temperature the dominant weight satisfies
\begin{equation}
\ln w(T_{p})=\ln g_{a,0}-\frac{\varepsilon_{a,0}}{T_{p}},
\end{equation}
substituting into \eqref{eq:S-Tp-apx} gives
\begin{equation}
\mathcal{S}(T_{p})=\ln g_{a,0}+\frac{\varepsilon_{b,0}-\varepsilon_{a,0}}{2T_{p}}.\label{eq:Stp-g's}
\end{equation}
The balance condition at $T_{p}$ yields
\begin{equation}
\frac{\varepsilon_{b,0}-\varepsilon_{a,0}}{T_{p}}=\ln g_{b,0}-\ln g_{a,0},\label{eq:Eg-conds}
\end{equation}
and substituting \eqref{eq:Eg-conds} into \eqref{eq:Stp-g's} leads
directly to
\begin{equation}
\mathcal{S}(T_{p})=\frac{1}{2}\bigl(\ln g_{a,0}+\ln g_{b,0}\bigr).
\end{equation}
This equals the interface residual entropy,
\begin{equation}
\mathcal{S}_{0}^{c}=\ln\sqrt{g_{a,0}g_{b,0}}.
\end{equation}
These results hold when the low-temperature behavior is governed by
two discrete sectors with $w_{0}\ll w_{a},w_{b}$ and when the limit
is taken along the pseudo-critical line $T_{p}(x)$; if additional
sectors become competitive or the spectrum is not gapped, the derivation
does not apply.

\end{document}